# C³IEL: Cluster for Cloud Evolution, ClImatE and Lightning

Daniel Rosenfeld[a*], Céline Cornet[b], Shmaryahu Aviad[c], Renaud Binet[d], Philippe Crebassol[d], Paolo Dandini[(2)], Eric Defer[e], Adrien Deschamps[d], Laetitia Fenouil[d], Alex Frid[f], Vadim Holodovsky[g], Avner Kaidar[f], Raphaël Peroni[b], Clémence Pierangelo[d], Colin Price[h], Didier Ricard[i], Yoav Schechner[g], Yoav Yair[j]

[a]The Hebrew University of Jerusalem; [b]LOA, Univ. Lille; [c]Israel Space Agency, [d]CNES, Toulouse; [e]Laero, CNRS/Univ. Toulouse, Toulouse; [f]Asher Space Research Institute, Technion, Haifa; [g]Viterbi Faculty of Electrical Eng., Technion, Haifa, [h]Departement of geosciences, Tel-Aviv University, Tel Aviv; [i]CNRM, Université de Toulouse, Météo-France, CNRS,Toulouse; [j]Reichman University, Herzliya
* Corresponding Author daniel.rosenfeld@mail.huji.ac.il

**Abstract**
Clouds play a major role in Earth's energy budget and hydrological cycle. Clouds dynamical structure and mixing with the ambient air have a large impact on their vertical mass and energy fluxes and on precipitation. Most of the cloud evolution and mixing occurs at scales smaller than presently observable from geostationary orbit, which is less than 1 km. A satellite mission is planned for bridging this gap, named "Cluster for Cloud evolution, ClImatE and Lightning" (C³IEL). The mission is a collaboration between the Israeli (ISA) and French (CNES) space agencies, which is presently at the end of its Phase A. The planned mission will be constituted of a constellation of 2 to 3 nano-satellites in a sun synchronous early afternoon polar orbit, which will take multi-stereoscopic images of the field of view during an overpass. C³IEL will carry 3 instruments: (1) CLOUD visible imager at a spatial resolution of 20 m. The multi-stereoscopic reconstruction of the evolution of cloud envelops at a resolution better than 100 m and velocity of few m/s will provide an unprecedented information on the clouds dynamics and evolution. (2) WATER VAPOR imagers at 3 wavebands with different vapor absorption will provide vertically integrated water vapor around the cloud and possibly a 3-dimensional structure of the vapor around the clouds due to their mixing and evaporation with the ambient air. (3) Lightning Optical Imagers and Photometers (LOIP). The lightning sensors will provide a link between cloud dynamics and electrification at higher spatial resolution than previously available. C³IEL will provide presently missing observational evidence for the role of clouds at sub-km scale in redistributing the energy and water in the atmosphere, and of the relation between storm vigor and frequency of lightning activity.

1. **Scientific background and main objectives**

The C³IEL mission is principally dedicated to the characterization of the convective clouds in the development stage. This includes the entire cloud species from marine stratocumulus, fair weather cumulus, cumulus congestus, the growing towers of cumulonimbus, thunderclouds, and severe convective storms.

Although the clouds encompass the scales from few hundred meters to hundreds of kilometers, they all share the fact that their growth occurs by a cascade of turbulent eddies that give the growing convective cloud elements their cauliflower appearance of an ensemble of perturbations which compose the cloud surface, as illustrated in Figure 1. Resolving and tracking these perturbations means resolving the cloud motions at which they are building and mixing with the ambient air. Larger cloud elements of several kilometers can be detected by current geostationary satellites, ground based radars can look at the dynamical properties of sufficiently thick cloud but the dynamical development of their small structures have never been resolved previously by satellite measurements. However, their quantification is essential for understanding the growth and organization of convective clouds (random organization or self-aggregation) and the way by which they influence the mean atmospheric state [1]. This is consequently a key element of the climate engine. The scale of the small convective elements that should be tracked is down to ~100 m. To resolve them clearly as identifiable features in the imagery, they have to be composed of several pixels. Therefore, a decametric spatial resolution at nadir is required. This resolution degrades with off nadir view but should remain high enough to capture the structure of small convective cloud elements of few hundreds of meters.

One of the main scientific objectives of the C³IEL mission concerns the dynamical properties of the cloud development. The methodology to retrieve the three components of cloud envelop development velocity relies on the mapping of the temporal evolution of cloud envelop topography at a resolution of several tens of meters. The derivation of the cloud envelop will be achieved by stereo-restitution using two or three simultaneous views of the cloudy scenes (Figure 2) from low orbit satellites. Visible high spatial resolution imagers on-board two or three satellites observing the same cloudy scenes several times during the satellite overpass will allow to achieve this stereo-retrieval and the derivation of the cloud development velocity.

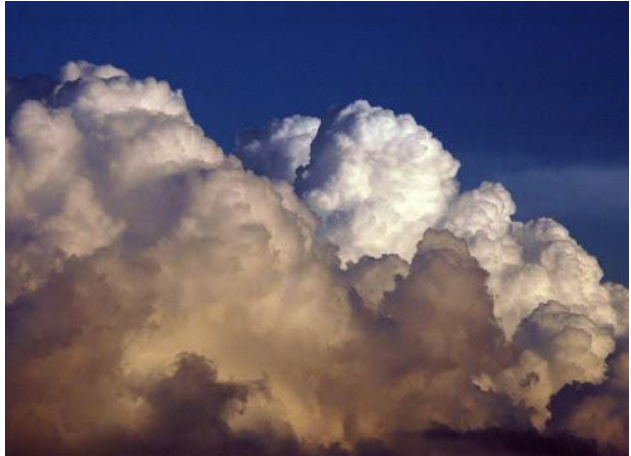

Fig. 1. Convective clouds. Each perturbation on the cloud surface represents a turbulent eddy at a scale of tens to few hundreds meters, which causes cloud mixing with and evaporation into the ambient air.

The formation and development of clouds in the atmosphere depends largely on the amount of water vapor available. In return, clouds, particularly those resulting from convective processes, contribute to redistributing water vapor in the atmosphere. Knowledge of the spatial and temporal variability of water vapor in a cloudy atmosphere is therefore essential information to improve our knowledge of the mechanisms linking water vapor and clouds. A stronger mixing with drier ambient air would inhibit more strongly the cloud development. Study of entrainment and detrainment near and above clouds requires a precise description of the water vapor fields around the cloud. This will be monitored by the water vapor content of the ambient air around the clouds which is highly variable. Therefore, the measurements of the cloud growth, turbulence and water vapor are highly synergistic. For the water vapor cameras, a high horizontal resolution is thus required in parallel with a good radiometric calibration to be able to capture small variabilities of the water vapor fields.

At a minimum, the integrated water vapor content next to and above the cloud can be achieved, but multi-angular information could be also used to obtain a coarse vertically resolved water content profile using tomographic methods

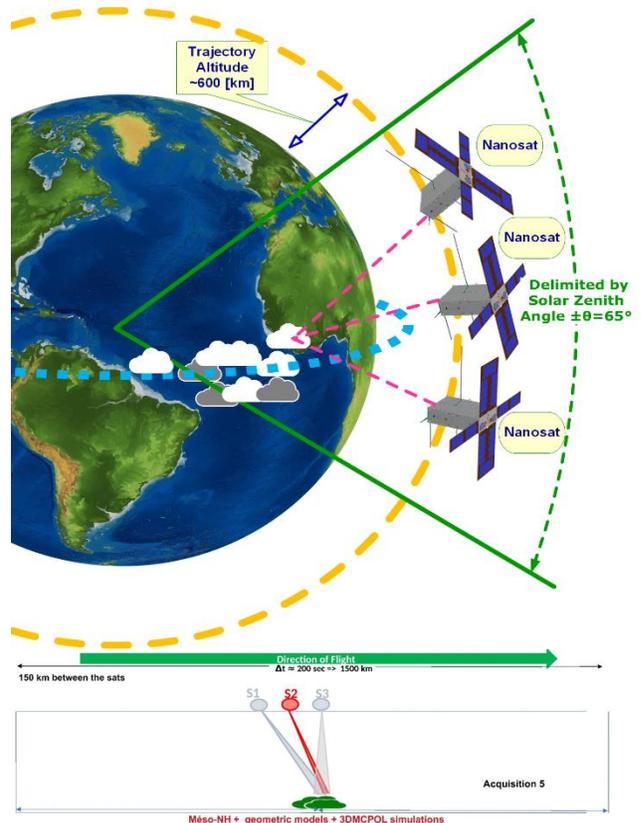

Fig. 2. Representation of the principle of measurements. Daytime acquisition of multi-stereoscopic images every 20 s during the 200 s overpass time, at a nadir resolution of 20 m for the CLOUD cameras.

The precipitation forming processes in the clouds lead to the development of downdrafts which are typically associated with the rain shafts. These downdrafts are the main cause that lead to the demise of individual convective rain clouds. However, when the downdrafts reach the surface they spread horizontally and can trigger gust fronts and the growth of new convective clouds. The combined aerosol cloud condensation nuclei and initial updrafts near cloud base determine cloud drop concentrations and sizes, which in turn determine the height at which the cloud will initiate its production of precipitation, which affect substantially the cloud dynamics and electrification [2], [3]. The $C^3IEL$ instruments will not allow to determine the number of cloud condensation nuclei (CCN). However, studies carried out in the case of cumulus clouds [4] and generalized to more developed convective clouds [5], show that the cloud drop effective radius is a function of its height from the cloud base and the number of activated condensation nuclei. Consequently, the knowledge of the effective radius at the cloud top lead to the number of activated condensation nuclei, which allows studies about aerosol-cloud interactions. This can be achieved using measurements of multispectral geostationary sensors (ABI, FCI) or at a better scale with multispectral radiometers on polar orbit such as the VIIRS/JPSS satellites. The important role of aerosols and the related cloud composition (cloud particle size, phase, number concentrations) will benefit that the measurements of C$^3$IEL will have several spatial and temporal coincidences with the JPSS polar orbiting satellites which do these aerosol and cloud measurements.

Collisions between hydrometeors during the development of the cloud induce electrification [6]. The electrical charges, mainly carried by the hydrometeors, induce an ambient electric field [7]. When the electrical field exceeds an altitude-dependent threshold [8], a lightning flash is triggered. This flash can stay in cloud (Intra-Cloud IC), or connect to the ground (Cloud-to-Ground CG). The lightning activity is then a by-product of complex microphysical, dynamics and electrification interactions. Indeed, the flash rate depends on vertical dynamics and microphysical movements within the cloud and can therefore be a good indicator of the severity of storms. Measuring simultaneously, and from the same platforms, the lightning activity, and the vertical and horizontal developments of parent convective clouds will help better refine the links between electrical, dynamics and microphysical properties competing within the clouds. As C$^3$IEL mission will be composed of nano-satellites, the lightning measurements will be conducted by a set of optical sensors, i.e. the Lightning Optical Imager and Photometers (LOIP). LOIP measurements will be available during both day and night, with a sufficient sensibility during daytime to extract the lightning optical signal within a rather bright cloud scene. During daytime, LOIP observational strategy, driven by the CLOUD observational requirements, will conduct continuous measurements during the 200-s for each scene sampled by CLOUD. During nighttime, LOIP observations will be conducted continuously along nadir. The lightning imager will provide a 2D mapping of the optical signal radiated at 777.4 nm within a few-nm band by the lightning discharges, signal scattered by the cloud hydrometeors and emanating from the cloud edges. The optical signal will be sampled at a time resolution of a few milliseconds and with a spatial resolution of a few hundred meters. The high temporal and spatial resolutions of the lightning imager should allow the mapping of the high-altitude discharge channels as demonstrated during the LSO (Lightning and Sprite Observations) experiment [9]. Lower altitude discharges will still be detected thanks to the scattering of the optical signal by the cloud hydrometeors, assuming that the sensor sensibility is high enough. The space and time mapping of the flashes will also provide some information on the inner structure of the clouds. On their side, the LOIP Photometers will measure the time evolution of the optical signal radiated by the lightning flashes within the entire field of view of the LOIP Imager during few hundreds of milliseconds at high temporal resolution. Several photometers operating at different wavelengths will document the lightning physical processes inducing light radiation in different spectral domains. Combined with simultaneous C$^3$IEL cloud observations, those multi-spectral lightning measurements should provide the required inputs to explore the properties of the atmospheric discharges (e.g. current flowing in the channel) through radiative transfer calculations and lightning modeling. In addition, the synergetic use of LOIP, CLOUD and the water vapor measurements should allow the study of the interactions between microphysical, dynamical, radiative and electrical processes and the verification of the numerical parametrizations implemented in cloud resolving models. Finally, the LOIP measurements on-board the C$^3$IEL mission are of major importance, specifically in high latitude areas that are poorly covered by geostationary lightning sensors, and where lightning flashes are expected to occur more often now with climate changes and to potentially trigger more wild fires (resulting in additional greenhouse gas emissions).

To summarize, the C$^3$IEL mission consists of a train of two to three identical synchronized nano-satellites to observe during daytime the same cloud scene at different observation angles. Each satellite will carry visible imagers (CLOUD) at decametric resolution, water vapor imagers (WVI), lightning imagers and photometers (LOIP). During daytime the lightning instruments will measure continuously while the CLOUD and water vapor imagers will obtain a succession of images of the same scene during the time of the satellites overpass. These unprecedented observations from space will document:

- Cloud top and 3D cloud envelop structures at a decametric resolution,
- Temporal evolution of cloud envelops, including cloud top vertical development velocity during the overpass time,
- Water vapor content around clouds,
- Electrical activity related to the convective processes that create the clouds,
- The links between lightning activity and upper tropospheric water vapor,
- Spatial organization of convective clouds with horizontal extent of few hundreds of meters.

## 2. Mission characteristics

C$^3$IEL mission is planned to be a two-year mission in a sun-synchronous polar orbit at an altitude of 600 km, with an Equatorial crossing time around 13:30.

Each day, the satellites will cover between 14 and 15 orbits around the globe. For each orbit, several daytime sequences of 200s of measurements will be performed for the CLOUD and WV imagers. Measurements will be realized every 10s to 20s during the time of overpass with the satellites pointing the same scene. A schematic representation of sequences of measurement is represented in Figure 2. As the CLOUD and Water Vapor imagers are passive sensors in the solar spectrum, measurements will be achieved uniquely during the day between 60°S and 60°N.

Concerning the CLOUD and WV imagers, the measurements, only during the day, consist of:
- Scenes: 2D arrays (images) of radiative quantities measured by the imagers on board one satellite: 1 image at 670nm (VIS) from the CLOUD imager, 3 images for the Water Vapor Imagers at 1.05, 1.13 and 1.38 µm abbreviated in the following as WV1, WV2 and WV3 respectively.
- Acquisition: Instantaneous measurements by the 2 or 3 nano-satellites. For VIS, WV1, WV2, WV3 channels an acquisition corresponds to 2 or 3 scenes per wavelength depending on the number of nano-satellites.
- Sequence: For VIS, WV1, WV2, WV3 channels, it corresponds to a series of 22 acquisitions (resp. 11) acquired every 10s (resp. 20s) during the time of the overpass with the nano-satellites pointing at the same scene during 200s. A sequence corresponds then to 11 to 22 acquisitions of 2 to 3 (depending of the number of nano-satellites) images per scenes for each imager (VIS, WV1, WV2, WV3 channels).

Concerning LOIP sensors, measurements will be acquired all along the orbit. Lightning measurements will indeed be conducted during the day in concordance with CLOUD and Water Vapor Imagers but also during the night.

LOIP imager measurements consist of successive frames of few milliseconds continuously recorded with identification of illuminated pixels. LOIP photometers acquisitions will consist of a series of spectral irradiance in function of time for several wavelength photometer acquisitions.

## 3. C³IEL Products

The Level 1 physical values will consist of reflectance corrected from instrumental and observational artefacts and geo-located in a 3D geographical space based on latitude and longitude.

### 3.1 CLOUD products

Level 1A: Sequence of geo-located images for each time of acquisition (satellite position) and each satellite with radiometric correction (N satellites x K positions per overpass).

Level 1B: Sequence of geo-located images for each time of acquisition (satellite position) and each satellite with geometric correction and geometric attributes (N satellites by K positions par overpass). Geographical coordinates (latitude, longitude) and observations geometries (solar zenithal and azimuthal angle, view zenithal and azimuthal angles)

Level 2A: Sequence of unregistered cloud top and cloud envelop points for each acquisition (satellite position), (K arrays per overpass).

Level 2B: Sequence of vertical cloud top development velocity and 3D motion vectors of cloud development (K-1 arrays per overpass).

### 3.2 Water Vapor products

Level 1A: Sequence of reflectance for each camera, each time of acquisition (satellite position) and each satellite with radiometric correction (3 channels x N satellites x K views per overpass).

Level 1B: Sequence of geo-located reflectance for each camera, each time of acquisition (satellite position) and each satellite with geometric correction and geometric attributes registered on CLOUD cameras in order to guarantee geometric consistency; Geographical coordinates (latitude, longitude) and observations geometries (solar zenithal and azimuthal angle, view zenithal and azimuthal angles).

Level 2A: Total column water vapor amount above cloud or ground surface.

Level 2B: Possibly also 3D water vapor structure, as an outcome of the tomography.

### 3.3 LOIP products

Level 1 for the lightning camera: 2D geo-referenced optical amplitude as a function of time, viewing times of each pixel or subset of pixels, optical amplitude thresholds used for each pixel or subset of pixels, background images and spacecraft attitude and flight parameters.

Level 1 for the photometer: calibrated time waveform of the optical signal with a geo-reference and spacecraft attitude and flight parameters.

Level 2A for the lightning camera: Validated events, groups, flashes, background scenes, viewing times. Optical pulses also called validated events will be merged in group, i.e. adjacent illuminated pixels in time and space, and groups in flash according to time and space criteria as already done for other instruments like LIS (Lightning Imaging Sensor) and GLM (Geostationary Lightning Mapper) (e.g. [10]). Background scenes could either be stored within the same files as the ones of event, group, flash dataset or within separate files. The viewing times will have to be stored in the L2A files as they will provide for each pixel or pixel subset of the imager the beginning and ending times of any location observed within the field of view of the imager.

Level 2B for the lightning camera: convective flag, flash density, flash rate at the scale of the electrical cells during viewing time, convective flag. Even if most users will use the L2A dataset, L2B dataset could also be generated to deliver an analysis of the lightning activity recorded within the field of view and during the period of observation.

Level 2 for the photometer: flag of flash occurrence (important for the cal/val and the verification of the imager), several parameters describing the waveforms (rise time, peak amplitude, duration).

*Acknowledgements*

The first author and presenter of this project (DR) is funded by the Israeli Ministry of Science and Technology grant number 2-17377.

C. Cornet, P. Dandini, E. Defer, R. Peroni and D. Ricard were supported by CNES.